# Memory-Efficient CNN Accelerator Based on Interlayer Feature Map Compression

Zhuang Shao, Xiaoliang Chen, Li Du, Lei Chen, Yuan Du, Wei Zhuang, Huadong Wei, Chenjia Xie, and Zhongfeng Wang, *Fellow, IEEE*

*Abstract*—**Existing deep convolutional neural networks (CNNs) generate massive interlayer feature data during network inference. To maintain real-time processing in embedded systems, large on-chip memory is required to buffer the interlayer feature maps. In this paper, we propose an efficient hardware accelerator with an interlayer feature compression technique to significantly reduce the required on-chip memory size and off-chip memory access bandwidth. The accelerator compresses interlayer feature maps through transforming the stored data into frequency domain using hardware-implemented 8×8 discrete cosine transform (DCT). The high-frequency components are removed after the DCT through quantization. Sparse matrix compression is utilized to further compress the interlayer feature maps. The on-chip memory allocation scheme is designed to support dynamic configuration of the feature map buffer size and scratch pad size according to different network-layer requirements. The hardware accelerator combines compression, decompression, and CNN acceleration into one computing stream, achieving minimal compressing and processing delay. A prototype accelerator is implemented on an FPGA platform and also synthesized in TSMC 28-nm COMS technology. It achieves 403GOPS peak throughput and 1.4x~3.3x interlayer feature map reduction by adding light hardware area overhead, making it a promising hardware accelerator for intelligent IoT devices.**

## I. INTRODUCTION

DEEP convolutional neural networks (CNNs) [1], compared to other traditional computer vision algorithms, offer significant accuracy improvement in target detection [2, 3], object recognition [4, 5], and video tracking [6-8]. However, state-of-the-art CNNs have become more complex and multi-branched [9] to achieve high prediction accuracy. During inferencing, CNN generates more than hundreds of megabytes of interlayer data and could significantly impact the hardware performance if the interlayer data is not properly stored [10]. Such impact could become even worse when CNNs are deployed on resource-limited mobile and IoT computing platform [11]. Limited by the on-chip memory size, many CNNs' interlayer feature maps are inevitably exchanged between on-chip and off-chip memories [12-14]. This not only results in large processing delay due to the data exchange but also dramatically increases the device power consumption as more than 70% of the entire system energy is wasted in the memory data exchange [15-17]. When processing high-resolution images, such as the satellite image in [18], the system delay and energy consumption caused by data exchange could be even larger. During external memory access, weights are priori information that can be pre-loaded during the process, and they only need to be transferred from the off-chip memory to the on-chip memory. Compared to weights, data is generated in real-time. Therefore, it is difficult to exchange data in parallel during the computation. What's more, data is exchanged in both directions, requiring twice bandwidth during access. Despite the processing delay and the power-constrained requirement, smaller memory size is always favorable to mobile and IoT devices as the on-chip memory size is directly related to the mass production cost, which is also very critical considering the wide application of mobile and IoT devices.

To solve the aforementioned concern, a hardware architecture for CNN inference acceleration with efficient interlayer feature maps storage and processing is necessary for CNN in mobile and IoT devices. Various hardware architectures are proposed in the previous works [19-30]. In [23], an efficient dataflow called "row stationary" was proposed which can provide high-level parallelism and data reuse. In [24], the power efficiency was improved in sparse CNN inference through powering off the computing units when the activation or weights are zero. However, these two works still store zero activation in on-chip memory, causing unnecessary resource wasted and lower efficiency. Some other works have explored methods in hardware design to reduce on-chip memory size and off-chip memory access [16, 25-29]. For example, in [25-28], only non-zero activations were processed and stored; the method is useful in scenarios with sparse feature maps. The sparsity of feature maps relies on the ReLU function that converts all negative activation to zero. However, some popular CNNs do not use ReLU as an activation function [31], resulting in very dense feature maps. The dense feature map will increase the overall storage overhead as the sparse-matrix compression requires additional index storage. In [29], a 4-bit activation quantization method was proposed, which improves memory

This work was supported in part by the National Natural Science Foundation of China under Grant 62004097 and 62004096, in part by the Natural Science Foundation of Jiangsu Province under Grant BK20200329.

Z. Shao, L. Du, X. Chen, Y. Du, H. Wei, C. Xie, and Z. Wang are with School of Electronic Science and Engineering, Nanjing University, China, 210023. W. Zhuang, L. Chen are with Beijing Microelectronics Technology Institute, Beijing, China, 100076. (Corresponding authors: Li Du and Yuan Du) storage and computation efficiency. However, the low-precision quantization also leads to severe CNN performance degradation [32]. In [16], a significance-aware transform-based codec was used to explore the correlation among feature maps, the low-correlation feature maps were regarded as the intrinsic representations of the original feature maps. The remaining feature maps were quantized and encoded to reduce external memory access. However, this technique was only used as an expansion of the existing accelerator to reduce off-chip memory access without optimizing the on-chip memory size.

Some works have used DCT for CNN training and original image compression to reduce storage space and transmission bandwidth. In [33], tensors of weights are transformed into the frequency domain by DCT for training and storage. In the inference, tensors with trained weights are transformed into filters of the same shape through IDCT for computation. Since some of the weights are set to zero in the frequency domain, there are fewer trained and stored weights than the original convolution layer. In [34], a DNN-favorable JPEG-based image compression framework called "DeepN-JPEG" is proposed. This framework can reduce the data storage and transfer overhead in IoT systems before transferring the real-time produced datasets. However, the above two works only compressed weights and the original images and did not solve the storage and data exchange problems caused by the massive interlayer feature maps.

Among all of the reported works, most of them did not compress the feature maps, causing significant on-chip storage overhead and off-chip memory access. In other works, either the compression of the interlayer feature maps was only fitted for one particular kind of CNN such as sparse CNN, or the compression process was not integrated with a CNN accelerator, which only improved off-chip memory access efficiency without reducing the on-chip memory size. To implement an architecture that can efficiently support arbitrary CNN interlayer data with different sparsity levels and on-the-fly compression process, four challenges must be explored:

1) The compression scheme needs to be efficient, achieving a high feature map compression ratio with light hardware overhead;

2) It should have wide applicability for arbitrary network structures;

3) The model accuracy loss due to the feature map compression should be negligible or within a tolerable range;

4) The hardware architecture has to combine hardware compression, decompression, and CNN acceleration together to achieve on-the-fly compression, reducing both on-chip memory size and off-chip memory access bandwidth.

In this paper, we propose a hardware architecture for CNN inference on mobile and IoT platforms. We focus on the compression of interlayer feature maps to achieve smaller on-chip memory size and less off-chip memory access. An 8x8 DCT is utilized to convert the feature maps to the frequency domain representation; quantization and sparse-matrix coding are adopted to compress the feature map. The hardware architecture can efficiently combine compression, decompression, and CNN acceleration into one computing stream, minimizing the processing delay caused by the compression module and achieving high throughput and parallelism. The main contributions of this paper are:

1) We adopt 8×8 DCT to perform frequency domain transformation on the interlayer feature maps with quantization to discard high-frequency components to compress the interlayer feature map size. The quantized interlayer feature maps are stored in the sparse matrix to further reduce the layer output size.

2) We design a CNN accelerator with an integrated DCT compression/decompression module. A novel data MUX in the processing element (PE) array is proposed to solve the overlapping problem caused by the convolution after 8×8 DCT.

3) We propose a reconfigurable on-chip memory scheme which can dynamically configure the feature map buffer size and the scratch pad size according to network layer requirements, resulting in less feature tiling and high PE utilization ratio for different kernel sizes.

4) We analyze the model accuracy loss due to removing the high-frequency components of the interlayer feature maps using DCT transform. Results show less than 1% accuracy loss due to DCT transform and compression.

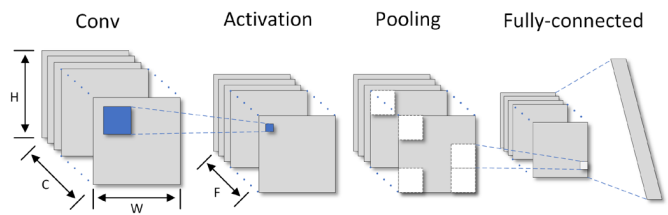

Fig. 1. The typical structure of CNNs.

The rest of this paper is organized as follows. In Section II, we introduce the basic concepts of CNNs and DCT. In Section III, the DCT-based compression and quantization scheme is discussed. Section IV describes the overall architecture of the accelerator. In Section V, the implementation details of each module are illustrated. The experimental results and the compression performance are given in Section VI. Section VII concludes this paper.

## II. BACKGROUND

### A. CNN Basic Conception

CNNs are mainly used for image and video processing. The architecture of CNNs consists of four typical layers for feature extraction and classification, including convolution layers, activation layers, pooling layers, and fully-connected layers. The output of each layer will be passed as the input of the next layer. The principal layer of the network is the convolutional layer, inserting the activation layer and the pooling layer in-between two convolutional layers can realize the nonlinear mapping and down-sampling of the feature maps. The fully connected layer is generally utilized as the last layer of the network to process the feature extraction results to obtain the final classification results. Fig. 1 shows the typical structure of CNNs.

*1) Convolutional Layer:* The convolution layer performs 2-

D convolution on the input feature maps. The filters and the input feature maps have the same number of channels. One filter and the input feature map are inner multiplied and accumulated on all channels to obtain one output pixel. A complete output feature map can be obtained by operating the filter to scan the entire input feature maps with a stride along with the filter height and width. Perform the same operation on different filters can obtain multiple output feature maps. The function expression of this algorithm is shown as (1).

$$O[f][r][c] = \sum_{ii=0}^{C} \sum_{i=0}^{K} \sum_{j=0}^{K} I[ii][r+i][c+j] \times W[f][ii][i][j]$$

(1)

In (1), C is the channel number of weight and input feature map, K is the weight kernel size.

*2) Activation layer:* The activation layer realizes the non-linear mapping between the convolutional layers. The most widely used activation functions include ReLU, sigmoid, tanh, etc. ReLU turns the negative number in the feature maps to zero, which provides significant sparsity for the feature maps. With the development of the network structure, leaky-ReLU, parametric-ReLU and Mish are introduced in the CNN [35], making the feature maps very dense.

*3) Pooling Layer:* The pooling layer aims at information extraction of adjacent pixels and down-sampling of feature maps, which achieves the elimination of redundant information and the expansion of the receptive field. Pooling layers are generally separated into two types: average pooling and max pooling. The average pooling calculates the average value of the pooling region, and the max pooling selects the maximum value of the pooling region.

### B. DCT Basic Conception

The discrete cosine transform (DCT) [36] is a mathematical operation relevant to the Fourier transform to analyze the frequency components of an image. In the Fourier series expansion, if the function to be expanded is a real even function, then the Fourier series contains only the cosine components, which can be discretized to derive the cosine transform. Therefore, the DCT can be considered as a finite sequence of data points represented by the sum of cosine functions oscillating at different frequencies. There are many forms of the DCT, the most commonly used pattern in image compression is the DCT-II, which is also the form used in this work. The expression of the 1-D DCT-II can be written as (2):

$$X_k = \sum_{n=0}^{N-1} x_n \cos\left[\frac{\pi}{N}\left(n + \frac{1}{2}\right)k\right] \quad k = 0, \dots, N-1 \quad (2)$$

where $x_n$ represent the origin signal, $X_k$ represent the coefficients after DCT transformation and N is the number of points of the original signal. Some algorithms multiply the first term $X_0$ by $1/\sqrt{2}$ and multiply the resulting matrix by the compensation coefficient $\sqrt{2}/\sqrt{N}$, making the DCT-II coefficients matrix orthogonal.

DCT-III, which is the inverse transform of DCT-II, is usually referred to as "IDCT" and its expression is shown as (3):

$$X_k = \frac{1}{2}x_0 + \sum_{n=1}^{N-1} x_n \cos\left[\frac{\pi}{N}n\left(k + \frac{1}{2}\right)\right] \quad k = 0, \dots, N-1 \quad (3)$$

The application of DCT on images is usually the two-dimensional transformation. The two-dimensional DCT of the image can be regarded as the decomposition of one-dimensional DCT along the rows and then along with the columns (or vice versa). Therefore, the two-dimensional DCT can be described as (4):

$$X_k = \sum_{n_1=0}^{N_1-1} \sum_{n_2=0}^{N_2-1} x_{n_1,n_2} \cos\left[\frac{\pi}{N_1}\left(n_1 + \frac{1}{2}\right)k_1\right] \cos\left[\frac{\pi}{N_2}\left(n_2 + \frac{1}{2}\right)k_2\right]$$

(4)

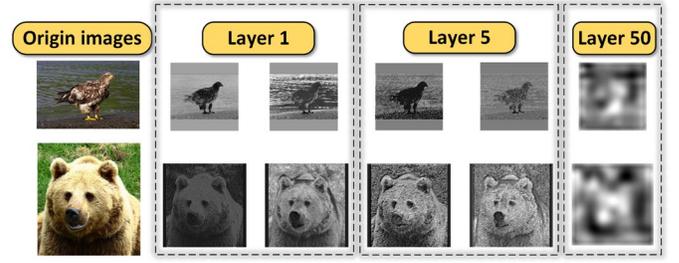

Fig. 2. Input images and interlayer feature maps of the Yolo-V3. Layer 1 and 5 shows the same object shape of the origin images while Layer 50 shows low-profile features.

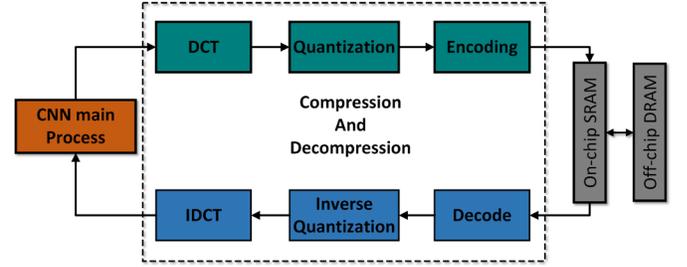

Fig. 3. Overall process of Compression and Decompression.

In general, when an image is transformed by 2-D DCT, most of its energy will be concentrated in the low-frequency components and its high-frequency components contain a lot of redundant information such as noise. This provides the chance for image compression in the frequency domain. Similar techniques are widely used in JPEG, HEIF MPEG, and other image and video standards [37].

## III. INTER-LAYER DATA COMPRESSION

### A. Motivation

The input of the CNN is usually an image with a fixed size. For a convolution layer, the output pixel is a linear mapping of the input pixel with the proper filter coefficients. Considering most of the operations in CNN are linear convolution, we can assume that the first few layers' output feature maps show similar image properties as regular input images. Fig. 2 is a comparison of the original image and the output feature maps in different layers. As it shows, the same object shape of the original image is also shown clearly in each feature map of the first few layers of the CNN such as Layer 1 and Layer 5. However, when the layer goes deeper, the original object becomes highly abstract feature information and does not show any input image object shape. Noticing this phenomenon, we consider applying DCT transform on each interlayer feature map of the first few CNN layers with quantization on the high-frequency components to compress its size. The overall procedure of the compression is shown in Fig. 3. After CNN main processing, each output feature map performs DCT transformation to get the frequency domain coefficients, the quantization and encoding are implemented sequentially to compress the feature maps before they are stored in the on-chip SRAM. Feature maps will be transferred to the off-chip DRAM if necessary. For the next convolution layer, the input feature maps are read from on-chip SRAM and decompressed for computation. Decompression is the inverse process of compression, decoder, inverse quantization and IDCT will be executed sequentially to reconstruct the origin feature maps. These processes will be introduced in III-B and III-C.

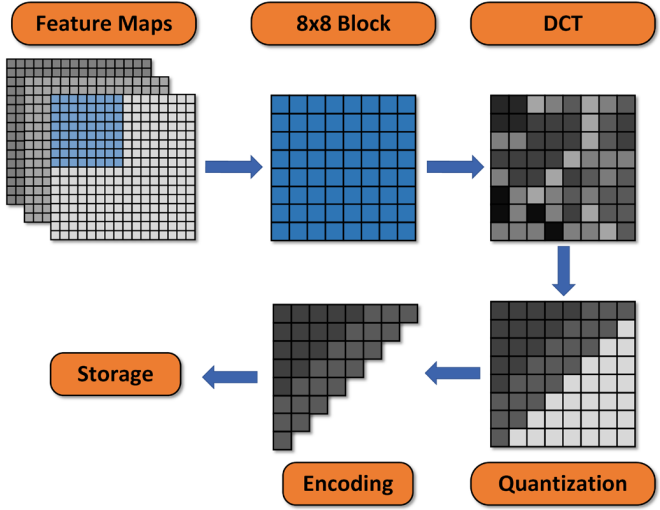

Fig. 4. Illustration of feature maps compression process

### B. Feature Maps Compression

Feature map compression is implemented through a DCT technology reported previously. It contains the following three parts: (1) an 8x8 kernel sized DCT to obtain feature map frequency information; (2) a quantization method with its corresponding quantization table (Q-table) to preserve low-frequency components of the feature while high-frequency components are quantized for compression; (3) an efficient encoding method to achieve high compression ratio and on-chip SRAM utilization. The feature maps compression process is visualized in Fig. 4:

*DCT:* Since the interlayer feature maps are often not very sparse in the image space, compressing them directly will introduce a large storage overhead of index and a low compression ratio. Therefore, DCT in the frequency domain is utilized. 8×8 block is selected as the basic unit for DCT because larger blocks bring more transform calculations and lead to higher computation delay and hardware overhead, while smaller blocks contain less boundary jumps so there will be less compression opportunity. The high-frequency components of the images correspond to redundant information such as noise and boundary jump. This feature permits us to discard high-frequency components to achieve the feature map compression with negligible accuracy loss. DCT is a natural and intuitive idea for obtaining different frequency domain components of the feature maps. The matrix form of 2-D DCT/IDCT can be written as (5) and (6):

$$Z = CXC^T \qquad (5)$$
$$X = C^TZC \qquad (6)$$

where $X$ is the input 8×8 matrix to be processed, $C$ is the 8×8 DCT transform matrix which can be derived from (4), $C^T$ is the transposition matrix of $C$, and $Z$ is 8×8 transformed result called DCT coefficients matrix. We use the 8×8 block as a basic unit for compression, decompression, convolution, and pooling, and define each 8 rows of feature maps as a row frame.

*Quantization:* Before encoding and compressing the DCT coefficients matrix $F_{freq}$, a quantization method is utilized to acquire more zeros. Our quantization method is divided into two steps: 1) low-precision General Matrix to Matrix Multiplication (GEMM). 2) Q-table quantization. Low-precision GEMM converts floating-point numbers into m-bit integers according to the range of the maximum and minimum values. Assuming that the maximum value in a given array is $F_{freq}^{max}$, the minimum value is $F_{freq}^{min}$, the maximum value of the quantized integer array is $i^{max}$ which equals to $2^m - 1$, and the minimum value is $i^{min}$ which equals to zero, then the Low-precision GEMM can be written as (7):

$$Q1_{freq} = round\left(\frac{F_{freq} - F_{freq}^{min}}{F_{freq}^{max} - F_{freq}^{min}} \times i^{max}\right) \qquad (7)$$

where $Q1_{freq}$ represents the quantized result of low-precision GEMM.

After completing the first step of quantization, we use the Q-table to quantize the DCT coefficient matrix. The design of the quantization table is based on the principle of preserving the low-frequency components and discarding the high-frequency components as much as possible. We refer to the JPEG Q-table which has small values in the top left part of the table and large values in the bottom right part of the table to achieve both high compression ratio and low accuracy loss. For different convolutional layers, the accuracy loss caused by lossy compression has different effects on the network performance. The first few layers' compression has negligible effect on network performance, while the medium layers' compression can result in noticeable performance degradation if the quantization loss is too large. To achieve minimal network accuracy loss, we configure a 2-bits register to achieve four different quantization levels between different layers. The selection of the quantization levels is through an off-line regression experiment on the test datasets. The Q-table values of the first few layers are larger in order to get a better compression ratio, while the Q-table values of the deep layers are adjusted to smaller values to ensure the accuracy of the network. The result of the Q-table quantization is as (8):

$$Q2_{freq}^{i,j} = round\left(\frac{Q1_{freq}^{i,j}}{Q_T^{i,j}}\right) \qquad (8)$$

where $Q2_{freq}$ represents the quantization result, $Q_T$ represents the 8×8 table and $(i, j)$ represents the coordinate of each element. The quantized $Q2_{freq}$ matrix has a large number of zeros in the matrix's bottom right corner, which is suitable for sparse encoding.

Correspondingly, the steps of inverse quantization are also divided into two steps to obtain the approximate matrix $F'_{freq}$ before quantization as shown in (9) and (10):

$$Q1'^{i,j}_{freq} = Q2^{i,j}_{freq} \times Q_T^{i,j} \qquad (9)$$

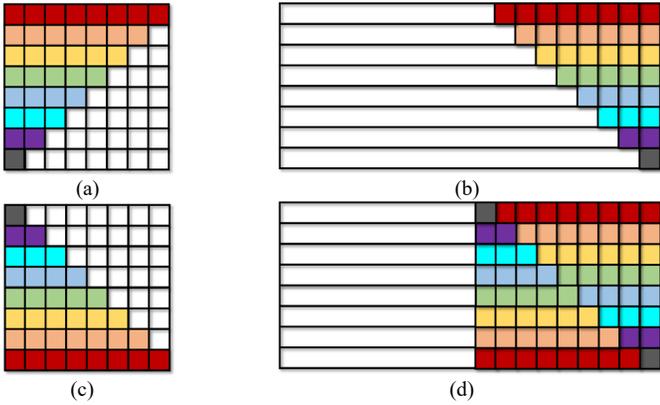

Fig. 5. Encoding and storage method, colored blocks represent the non-zero data while white blocks represent the zeros. (a) an even number 8×8 quantized matrix data distribution, and (b) the state of the SRAM after storing the first 8×8 matrix. (c) an odd number 8×8 quantized matrix after flipping, and (d) the state of the SRAM after storing the second 8×8 matrix.

$$F'_{freq} = \frac{Q1'^{i,j}_{freq}}{i^{max}} \times (F^{max}_{freq} - F^{min}_{freq}) + F^{min}_{freq} \qquad (10)$$

where $Q1'^{i,j}_{freq}$ represents the feature maps after the first step of quantization, $F'_{freq}$ represents the origin feature maps.

***Encoding:*** To compress the final storage size, a proper encoding method that can exploit the sparsity of the matrix and maintain a regular memory storage pattern is required as one feature map is divided into multiple 8x8 transformed matrixes that will be stored in one shared memory. In addition, the encoding method is also required to support fast hardware encoder and decoder implementations with smooth memory storage and read out data flow, thus the processor throughput will not be affected due to data congestion in the encoding or decoding module.

Ideally, the 8×8 quantized matrix can be regarded as a one-dimensional vector in a zig-zag sequence and Huffman coding is the best method to achieve the theoretical highest compression ratio. However, the implementation of Huffman encoding and decoding will request a look-up table which introduces considerable hardware overhead. In addition, the variable code length in Huffman coding results in an irregular size encoded symbol. During decoding, the next symbol place cannot be properly determined in the code sequence until the current symbol is fully decoded. Thus, symbols cannot be decoded in parallel, which limits the overall hardware decoding speed. Therefore, we propose a simple sparse matrix encoding method which adapts the quantized matrix sparsity and provides a fast-encoding scheme that can support on-the-fly data compression.

For a specific quantized matrix, we only store non-zero data and discard all zero data. A 1-bit 8×8 index matrix is used to indicate whether the data under a certain coordinate in the original 8×8 quantized matrix is zero, where the index matrix value 1 indicates that the data in the corresponding position of quantization matrix is non-zero, otherwise it is zero. When reading or writing data to the buffer, the index matrix can be utilized to determine whether to access the corresponding SRAM. At the same time, it can be used as the gate signal of the multiplier in the IDCT module to skip IDCT matrix calculation. If the index is 0, the multiplier is turned off to save power.

The index matrix is sequentially stored in a dedicated SRAM called index buffer. The quantized matrix is stored in the feature map buffer. The storage method is shown in Fig. 5. The feature map buffer consists of 8 pieces of SRAM and each SRAM is used to store one row of data. Based on the above analysis, most of the zeros are located at the bottom right corner of the matrix. As described in Fig. 5. (a), the colored blocks in the 8×8 quantized matrix are non-zero data while the white blocks indicate zeros. The non-zero data in the even number quantized matrices is directly stored in SRAM by column and the state after the storage is shown in Fig. 5. (b). If the subsequent quantized matrices are stored in this way, there will be a lot of vacancy in the last row of SRAM while the first row of SRAM is full, resulting in low SRAM utilization. To improve the memory utilization, the next sequential 8x8 quantized matrix data will be flipped and stored in the memory with a reverse order as shown in Fig.5 (c), (d). The current matrix's row 8 will be stored together with the previous matrix's row 1. Through the matrix flipping between two sequential matrixes, the overall memory utilization is improved.

Reading data from the feature map buffer is the inverse process of the above method. According to the index matrix, the chip select signal and address of the SRAM can be determined to get the non-zero data then padding the zero data in the corresponding position. When processing the next matrix, the readout data will be inverted.

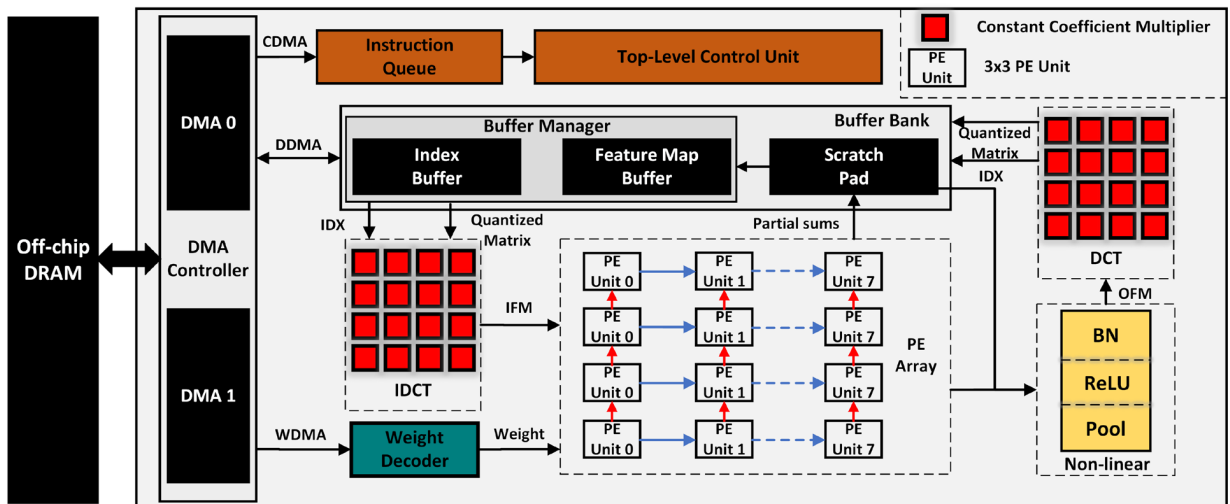

Fig. 6. The overall hardware architecture.

### C. Decompression

Before the convolution of the next layer, the quantized matrix stored in the feature map buffer needs to be read out for decompression to restore the approximate feature maps. The decompression process consists of three steps: decoding, inverse quantization, and IDCT.

Decoding is executed by reading a set of non-zero data from the feature map buffer according to the index matrix and restore it to the corresponding position in the quantized matrix. According to (9), (10) in III-B, the data is inverse quantized by the Q-table to obtain the DCT coefficients matrix. Finally, the approximate feature maps can be obtained by matrix multiplication in the IDCT module with the DCT coefficients matrix.

## IV. SYSTEM ARCHITECTURE

To apply the interlayer feature maps compression technique, a CNN accelerator with integrated compression and decompression module is proposed. The overall architecture is shown in Fig. 6. The input image, weights, and system instructions are fetched from the external memory via a direct memory access (DMA) controller to the on-chip memory. The DMA controller has two sub-modules, one for transmitting feature maps and the other one for transmitting weights in parallel. 16 bits dynamic fixed-point data format is adopted in this design to obtain comparable accuracy to float 32 bits [38]. The models are 8-bit feature-wise quantized [39] for inference. The feature maps are stored into the feature map buffer, the accelerator instructions are stored in the instruction queue for parsing and execution, and the weights are decoded by the weight decoder with a preload buffer in the PE array to calculate convolution.

The accelerator incorporates a 480 KB single-port-SRAM called buffer bank which is divided into three parts: feature map buffer, scratch pad, and index buffer, to store sparse matrices of the compressed interlayer feature maps, partial sums, and sparse matrix indexes respectively. The size of feature map buffer and scratch pad can be dynamically configured according to the network layer requirements. The feature map buffer is divided into two segments, buffer A and buffer B as a ping-pong-buffer. A buffer manager is used to manage input and output I/O connections from this SRAM through MUXs. The initial size of these two buffers is 128 KB, one is utilized for the input feature maps and the other is for the output feature maps of the current layer. Index buffer is 32 KB and is also divided into two parts, which are used to store the sparse matrix indexes of the corresponding sparse matrices stored in feature map buffer. Since the zeros are concentrated in the bottom part of the quantized matrix, a flip method is adopted for the storage to maximize the on-chip memory utilization. The instruction queue fetches the instructions and loads them into the local memory when the accelerator is enabled, then these instructions will be executed in order. The top-level control unit manages all the registers required by the accelerator to configure each sub-module. The DCT module consists of a DCT process unit and an IDCT process unit. Each unit contains 128 constant-coefficient multipliers to process 4 channels' DCT/IDCT in parallel and achieves pipelining with convolution. The PE array contains 288 PEs to achieve high parallelism. The PE array can support up to 4 input channels and 8 rows for 3×3 convolution in parallel. In 1x1 convolution, one PE will be turned off due to the limited data bandwidth. Weight decoder periodically pre-loads weights from external memory through DMA with a FIFO to hide the access latency. The partial sums calculated by the PE array are sent to the scratch pad for further accumulation and storage. The initial size of the scratch pad is 64 KB and can be configured up to 192 KB dynamically. A non-linear module is designed to process non-convolutional operations such as ReLU, batch normalization (BN), or pooling functions.

The accelerator starts to perform the convolution when the instruction queue executes the CONV instruction. First, the IDCT module will fetch the compressed sparse matrices and the sparse matrix indexes from the buffer bank for inverse quantization and IDCT to reconstruct the approximate feature maps, then they will be sent to the PE array for convolution. The scratch pad will accept the partial sums sent by the PE array to accumulate and store. After all the input channels are calculated, the results will be sent to the non-linear module to execute non-convolutional operations, and then the output feature maps will be passed to the DCT module for compression and store back to the feature map buffer.

## V. DATA FLOW AND MODULE IMPLEMENTATION

To better combine the compression process and the convolution computation, we propose an efficient data flow and module implementation in this article. This section will discuss them in details.

### A. Data Flow and Reuse Scheme

The overall computation flow follows the weight reuse strategy. The weight of a specific input channel and output channel is stored in the accelerator before scanning through the entire input feature maps. Benefitting from our compression method, the input feature maps can be stored on-chip without external memory access to DRAM in most cases. On-chip stored weights will be transferred only when they need to be updated; there is no need to read the off-chip stored weight repeatedly.

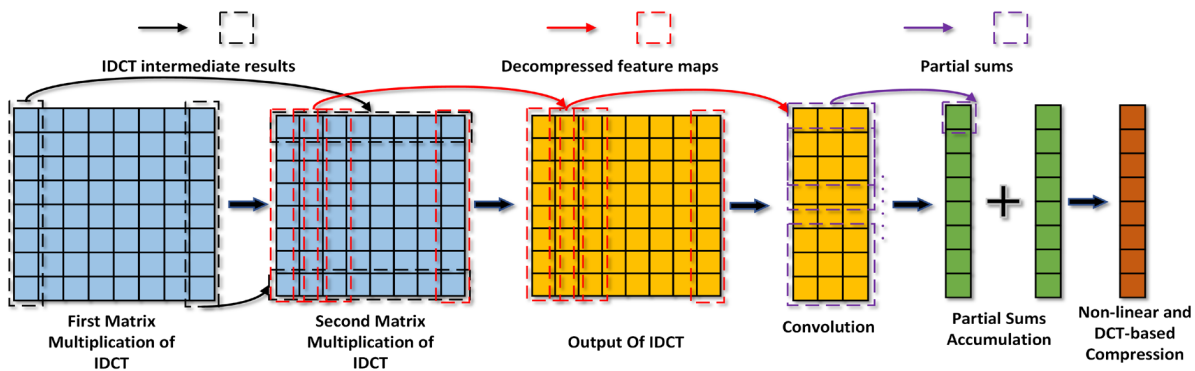

Fig.7. An Example of the computation flow.

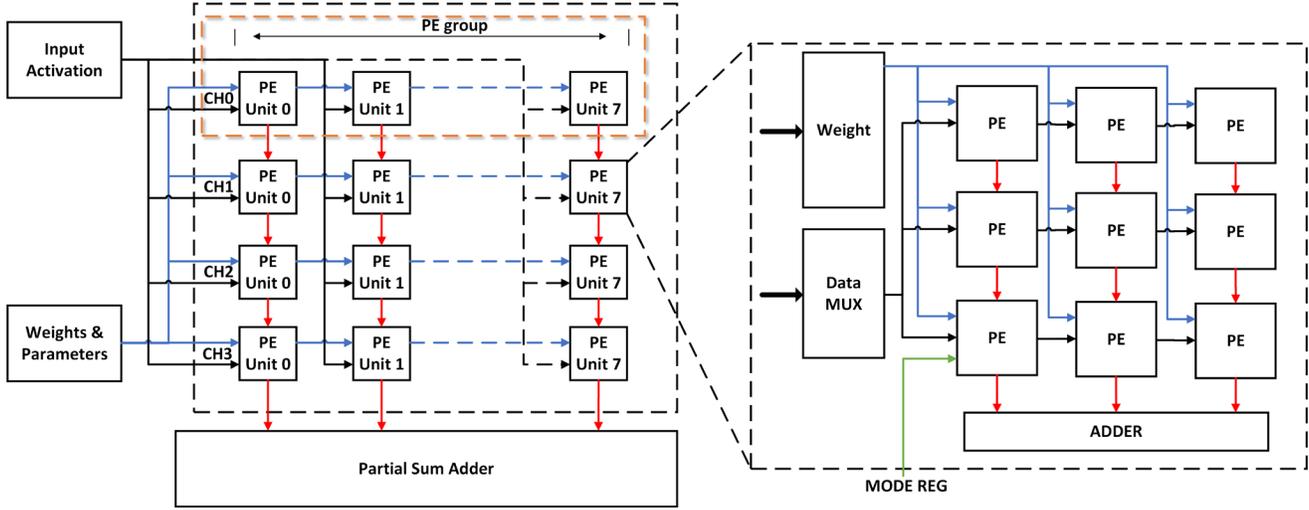

Fig. 8. Overall architecture of the PE array.

Four output features are calculated simultaneously to avoid repeated decompression of the input feature maps in 3×3 convolution mode. Specifically, we adopt four cycles to calculate 4 output feature maps corresponding to the same input feature maps and four different convolution filters, and the PE array will automatically switch the weights of different filters in each cycle. Four input feature maps will be fed into the PE array together to ensure that the input and output bandwidth is consistent. Weights will be pre-load to a local buffer embedded in the PE array before the start of the convolution, and the weight for the next calculation will be pre-read during the convolution. The bandwidth between calculation modules such as PE array, non-linear, and DCT/IDCT is eight rows by one column.

At the beginning of the convolution, the IDCT module will read a column of the quantized matrix from the feature map buffer according to the index matrix to perform the inverse quantization. The calculation of the IDCT involves two cosine function multiplication as (4) and is divided into two steps. The IDCT will finish the first cosine function multiplication after reading eight columns of data, then sequentially begin to calculate the second matrix multiplication to get the input feature maps of the current layer. The input feature map will be sent to the PE array for convolution. The PE array begins 3x3 kernel computation after receiving three columns of feature maps. Partial sums are accumulated and stored in the scratch pad. The final output feature maps after partial sums accumulation are sent to the non-linear module and DCT module to perform the non-linear operation and DCT compression process. Fig. 7 illustrates an example of this process in detail.

During 1×1 convolution, we calculate 8 output feature maps with 8 filters in parallel to achieve input reuse. Unlike 3×3, in 1×1 convolution mode one PE will be turned off to save power, and the remaining 8/9 PEs will calculate 8 output feature maps with 8 filters in one clock cycle. This will cause the input data bandwidth of the scratch pad to be twice of the 3×3 convolution mode. In the V-C chapter, we will introduce a reconfigurable on-chip memory scheme to solve this problem.

The kernel size supported by the accelerator is up to 7×7, and the supported stride of the convolution can be 1 or 2. For convolution with a kernel size larger than 3×3, a filter decomposition technique reported in [14] is used to decompose the larger kernel size into multiple 3×3 kernels for computation. If the stride of convolution is two, the PE array will use one clock cycle to bypass the columns that do not need to be computed. The validity of the PE array output rows will be indicated by the row valid signals.

### B. PE Array

The PE array contains 32 PE units, each of 8 PE units generates a PE group for calculating one input feature map, and a total of four input feature maps can be calculated in parallel. Each PE unit contains 9 multipliers and adders to support convolution up to 3×3.

The description of the PE array is shown in Fig. 8. Input feature maps of eight rows and four channels are sent to four PE groups respectively. In a PE group, eight rows of data are fed to the data MUX while 3 rows are selected and sent to the corresponding PE units. The data transmission between PEs is realized through D flip-flops, the data in each PE will be sent to the right neighbor PE when a new column is obtained. The weights are stored in a local buffer and delivered to each PE unit. The MODE register will be used to turn off unused PEs according to the convolution mode. The multiplication results of PEs are accumulated in the PE units first, then eight partial sums are obtained in the PE array through channel accumulation to reduce the frequency of the scratch pad accessing.

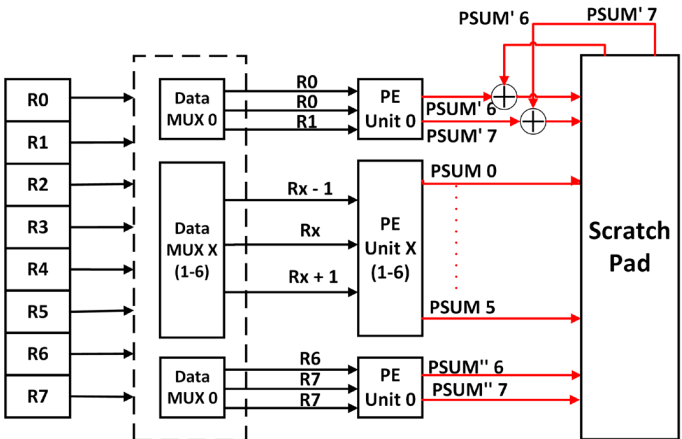

Fig. 9. Design of data MUX and partial sums accumulation scheme.

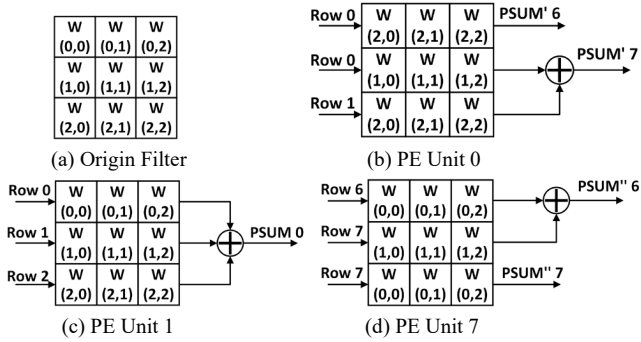

Fig. 10. Convolution pattern of PE units.

Since we select 8×8 blocks for compression and convolution, the overlapping problem of adjacent 8 rows called row frame (RF) occurs in 3×3 convolution mode. The typical method that stores the overlapping part and reads them when calculating the next RF, creates significant memory overhead. To solve this overlapping problem, we design a data MUX module for 8 PE units to select different data and weights. Fig. 9 illustrates the three rows of data selected by data MUX and the connection between the corresponding PE units and the scratch pad. In Fig. 9, PSUM represents the partial sums of the current RF, PSUM′ represents the partial sums of the previous RF and PSUM′′ represents the partial sums of the next RF. Fig. 10 shows all of the possible connections between the data, weights, and outputs to the 8 PE units. The calculation pattern of PE unit 1 to PE unit 6 is shown in Fig. 10. (c). These six PE units calculate $PSUM_0$ to $PSUM_5$ respectively. These six partial sums are the "completed partial sums" calculated by the original filter, which only need to be accumulated in the channel direction and do not need to be spliced with the partial sums of the next RF. PE unit 0 calculates $PSUM_6'$ and $PSUM_7'$, which need to be accumulated with the previous RF partial sums saved in the scratch pad. The accumulation results are stored back to the scratch pad. PE unit 7 calculates $PSUM_6''$ and $PSUM_7''$, which need to be accumulated with the next RF. Therefore, the partial sum results are stored in the scratch pad directly, waiting for the accumulation of the next RF to obtain the "complete partial sums". The weights and data required for PE unit 0 and PE unit 7 are shown in Fig. 10. (b) (d), respectively. PSUM′ and PSUM′′ can be accumulated simultaneously with channels partial sums accumulation without causing additional overhead.

### C. Memory Configuration and Non-linear Module

The overall block diagram of on-chip memory and non-linear modules is shown in Fig. 11. The buffer bank contains a feature map buffer and a scratch pad to store feature maps and partial sums. The non-linear module can process BN, ReLU and pooling in different sequences.

As described in Section V. A, parallel output feature maps are adopted to reduce the frequency of IDCT module computations. However, this also brings a four times partial sum storage overhead compared to a single feature output. In addition, the neural network has a typical feature that when the layer is deeper, the size of the feature map gradually decreases, but the number of channels gradually increases, which leads to an increase in the demand for the size of the feature map buffer, and the scratch pad tends to have a large free unused space. Based on the above observation, a large scratch pad size is more favorable to the first few layers. However, it will result in a low memory utilization ratio if the scratch pad size cannot be changed.

To solve the above conflict, a reconfigurable on-chip memory scheme is proposed to dynamically configure the feature map buffer size and scratch pad size. The overall memory architecture is shown in Fig. 11. The scratch pad is dedicated to PSUM accumulation and its size is 64 KB. The feature map buffer A and B are used to store compressed layer input and output data. Each of the feature map buffer is 128 KB and paired with a configurable memory. Each configurable memory is set to 64 KB including two sub-banks, and each sub-bank is 32 KB. Therefore, the scratch pad can be configured to 64 KB, 128 KB or 192 KB, and each feature map buffer can be configured to 128 KB, 160 KB or 192 KB. Depending on the layer requirement, each of the configurable memory banks can be either combined into feature map buffer or scratch pad, extending their depths dynamically.

In 3x3 convolution mode, 10 rows and 4 channels partial sums will be sent to the scratch pad each time, where 8 rows are the partial sums of the current RF, and 2 rows are the partial sums of the next RF. In 1x1 convolution mode, there is no overlapping problem in convolution, the scratch pad will be used to store 8 channels partial sums in parallel, which can increase the PE utilization to 8/9 in 1x1 convolution mode.

When the partial sums of the last channels are accumulated, the scratch pad will transmit the results to the non-linear module to perform non-linear computations such as ReLU, BN, and pooling based on the network structure. The coefficients required by BN are extracted during training and transmitted to the accelerator together with weights during inference. The computation sequence of the ReLU, BN, and pooling in the non-linear module can be configured to fit for different kinds of CNN applications. Finally, the output of the non-linear module will be sent to the DCT module for further compression before storing back to the memory.

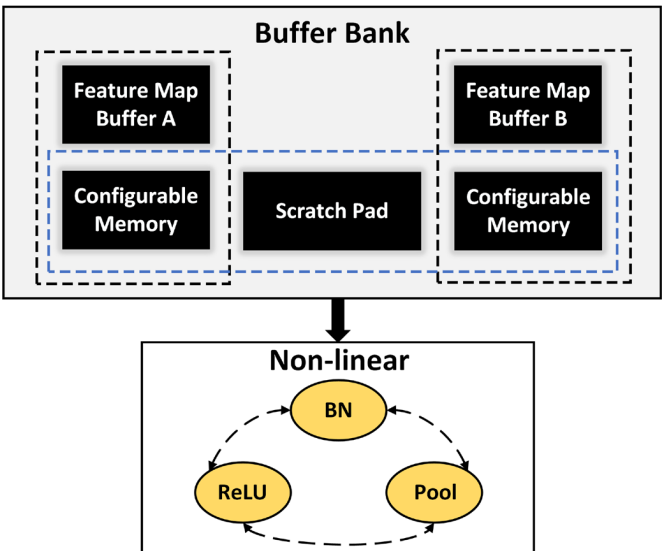

Fig. 11. SRAM configuration and non-linear module implementation.

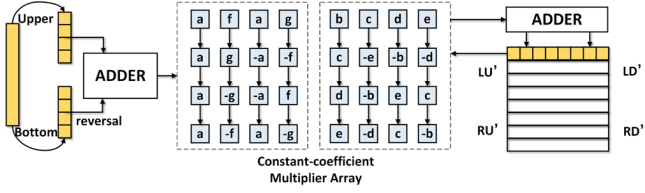

Fig. 12. DCT module implementation.

### D. DCT/IDCT

There have been many works to achieve efficient hardware implementations of DCT/IDCT. In this paper, we refer to a fast DCT/IDCT algorithm [40] and integrate it into our accelerator. This work disassembles one 8×8 matrix multiplication into four 4x4 matrix multiplications; It benefits from the particular characteristics of the DCT matrix, saving half of the computing resources.

As described in [40] and (5), we can decompose the input matrix $X$ and transform matrix $C$ as (12) and (13):

$$X = \begin{bmatrix} X_{LU} & X_{RU} \\ X_{LD} & X_{RD} \end{bmatrix} \quad (12)$$

$$C = Q^T \begin{bmatrix} Ce & CeP \\ Co & CoP \end{bmatrix} \quad (13)$$

where $Q$ and $P$ are shown in (14):

$$Q = \begin{bmatrix} 1 & 0 & 0 & 0 & 0 & 0 & 0 & 0 \\ 0 & 0 & 1 & 0 & 0 & 0 & 0 & 0 \\ 0 & 0 & 0 & 0 & 1 & 0 & 0 & 0 \\ 0 & 0 & 0 & 0 & 0 & 0 & 1 & 0 \\ 0 & 1 & 0 & 0 & 0 & 0 & 0 & 0 \\ 0 & 0 & 0 & 1 & 0 & 0 & 0 & 0 \\ 0 & 0 & 0 & 0 & 0 & 1 & 0 & 0 \\ 0 & 0 & 0 & 0 & 0 & 0 & 0 & 1 \end{bmatrix}, \quad P = \begin{bmatrix} 0 & 0 & 0 & 1 \\ 0 & 0 & 1 & 0 \\ 0 & 1 & 0 & 0 \\ 1 & 0 & 0 & 0 \end{bmatrix} \quad (14)$$

$Ce$, $Co$ are matrix transforming coefficients get from (4), as shown in (15):

$$Ce = \begin{bmatrix} a & a & a & a \\ f & g & -g & -f \\ a & -a & -a & a \\ g & -f & f & -g \end{bmatrix}, \quad Co = \begin{bmatrix} b & c & d & e \\ c & -e & -b & -d \\ d & -b & e & c \\ e & -d & c & -b \end{bmatrix} \quad (15)$$

Further, we can get the $Z'$ from (16) and (17):

$$Z' = QZQ^T = \begin{bmatrix} (Y'_{LU} + Y'_{LD}P)Ce^T & (Y'_{LU} - Y'_{LD}P)Co \\ (Y'_{RU} + Y'_{RD}P)Ce^T & (Y'_{RU} - Y'_{RD}P)Co \end{bmatrix} \quad (16)$$

$$Y'_{LU} = Ce(X_{LU} + PX_{LD})$$
$$Y'_{LD} = Ce(X_{RU} + PX_{RD})$$
$$Y'_{RU} = Co(X_{LU} - PX_{LD})$$
$$Y'_{RD} = Co(X_{RU} - PX_{RD}) \quad (17)$$

Finally, transformed DCT coefficients $Z$ can be obtained by (18):

$$Z = Q^T Z' Q \quad (18)$$

We designed a hardware architecture to implement the above algorithms. As shown in Fig. 12, there is a constant-coefficient multiplier (CCM) array containing 128 CCMs to perform DCT matrix multiplication. Every 32 CCMs can complete the multiplication of the 8×8 matrix and the 8×1 matrix in one cycle to obtain an 8×1 matrix, thus four channels can be calculated in parallel, saving half of the computing resources. The input 8×1 matrix is divided into upper and bottom parts, the bottom part will be reversed at first, then added to the upper part and sent to the CCM array as input. The CCM array will output an 8×1 matrix after calculation to the adder for reversal and accumulation, and the result of the accumulation will be sent to the CCM array again for 1×8 matrix and 8×8 matrix multiplication to obtain a final 1×8 matrix result.

TABLE I
HARDWARE SPECIFICATIONS

| Technology | TSMC 28nm |
|---|---|
| Clock Rate | 700 MHz |
| Gate Count | 1127 K |
| Dynamic Power Consumption | 186.6 mW |
| Core Area | 1.65mm × 1.3mm |
| Number of PEs | 288 |
| On-chip SRAM | 480 KB |
| Index Buffer | 32 KB |
| Feature Map Buffer | 256~384 KB |
| Scratch Pad | 64~192 KB |
| Supply Voltage | 0.72 V |
| Peak Throughput | 403 GOPS |
| Arithmetic Precision | 16-bit fixed-point |
| Energy Efficiency | 2.16 TOPS/W |
| Supported CNN operations | Conv 1x1-7x7 |
| | Depth wise Conv 1x1-7x7 |
| | Batch Normalization |
| | ReLU |
| | Leaky ReLU |
| | Program ReLU |
| | Pooling |
| Number of CCMs in DCT Module | 128 |
| Number of CCMs in IDCT Module | 128 |

## VI. RESULT

In this section, we reported the hardware performance of the accelerator and evaluated a detailed test on the compression performance with five state-of-the-art CNNs: VGG-16-BN, ResNet-50, MobileNet-v1, MobileNet-v2, and Yolo-v3. We analyzed the data distribution, compression ratio, and compressed data distribution of each layer of these CNNs. The results proved that our method can effectively reduce the required on-chip storage size and off-chip memory access with less than 1% accuracy loss.

### A. Hardware Performance

To evaluate the hardware performance, we implemented the accelerator using Synopsys Design Compiler (DC) with the TSMC 28nm HPC+ technology library from the ss (slow-PMOS and slow-NMOS) corner, the supply voltage is 0.72 V, and the temperature is 125 °C. The layout characteristics of the accelerator are shown in Fig. 13, and the core dimension is 1.65mm × 1.3mm. The detailed specifications are listed in Table I. The peak throughput is 403 GOPS with a 700 MHz clock frequency and the dynamic power consumption is 186.6 mW, corresponding to 2.16 TOPS/W in energy efficiency. The dynamic power of the accelerator is estimated by Synopsys PrimeTime PX using VGG-16-BN as the benchmark. The accelerator supports most CNN operators such as convolution, ReLU, BN, and Pooling. We also implemented it on Xilinx Zynq XC7Z045FFG900-2. The LUT, FF, DSP, and BRAM resource utilization are 110K(51%), 39K(9%), 544(60%), 129(24%) respectively with a 50 MHz clock frequency.

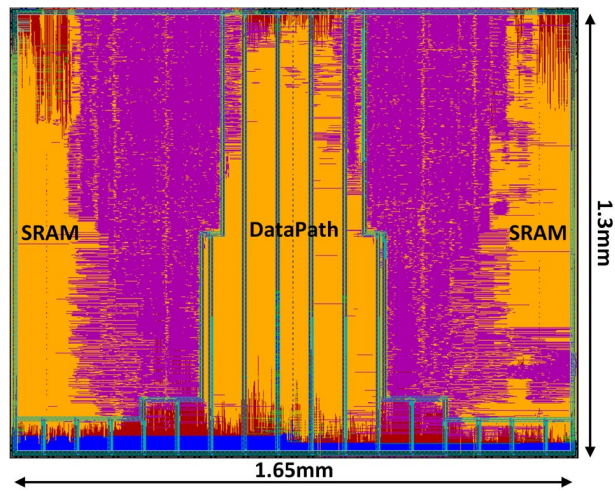

Fig. 13. Layout view of the accelerator.

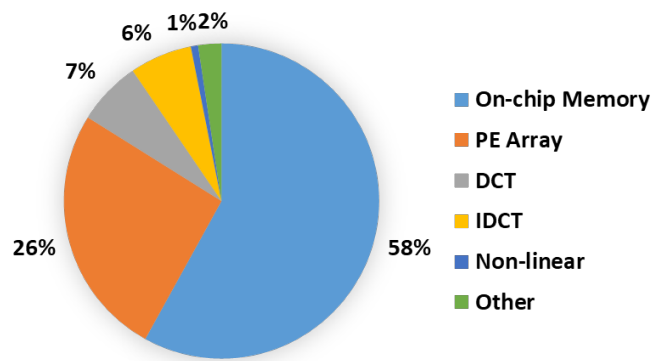

Fig. 14. Area breakdown of the accelerator.

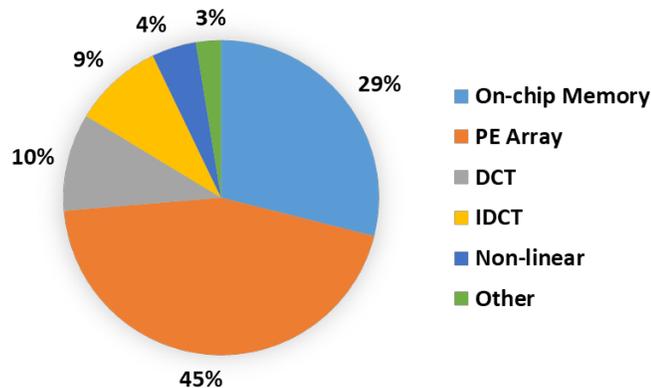

Fig. 15. Power consumption breakdown of the accelerator.

The overall gate count of the accelerator, excluding on-chip SRAMs is 1127K NAND-2 gates. Fig. 14 shows the area breakdown and Fig. 15 shows the power consumption breakdown of the accelerator. The total SRAM size of this accelerator is 480KB. The area of the SRAMs takes over half of the total area. The SRAMs are categorized into feature map buffer and scratch pad. The size of the feature map buffer can be configured from 256KB to 384KB, while the scratch pad can be configured from 64KB to 192KB. Since the dimension of the feature map is large in the first few layers, the feature map

TABLE II
EXTERNAL MEMORY ACCESS SAVED BY THE COMPRESSION METHOD

| Network | Data Reduction (MB/Fig) | Time Reduction* (ms/Fig) | Power Overhead** (mW) | Power Reduction** (mW) |
|---|---|---|---|---|
| Yolo-v3 | 54.36 | 14.12 | 6.9 | 117.8 |
| ResNet-50 | 33.10 | 8.56 | 15.1 | 555.2 |
| VGG-16-BN | 26.44 | 6.87 | 35.8 | 155.9 |
| MobileNet-v1 | 18.11 | 4.70 | 15.7 | 2592.9 |
| MobileNet-v2 | 20.19 | 5.24 | 11.4 | 4009.4 |

*The speed of data access refers to the DMA module supported by SYNOPSYS DW-axi-dmac-A415-0. (*/Fig) means data or time reduction per inference image.
**The power overhead of the DCT/IDCT modules and the power reduction of the off-chip memory access. The average off-chip memory access energy is 70pJ/bit.

buffer is configured as 256KB in most cases. The total size of the SRAMs, feature map buffer and scratch pad are selected based on the trade-off between the area, performance, and the compressed feature map size. The PE array consists of 288 PEs and occupies 26% of the total area. DCT and IDCT include their hardware modules such as quantization, encoding, and constant-coefficient multipliers. Due to the small area size of the constant-coefficient multiplier, the additional overhead brought by the interlayer feature map compression is only 13% of the area of the accelerator. The DCT/IDCT modules introduce 19% of dynamic power consumption, and for the layers that do not require compression, these modules are turned off to save dynamic power consumption. Since the DRAM access is the most energy consuming data movement per access [23], the reduction of on-chip/off-chip memory access bandwidth saved by compression can also reduce power consumption substantially. Compared with the power consumption introduced by the DCT/IDCT module, the overall power consumption is still reduced. The detailed power improvement is shown in Table II.

### B. Compression Performance

Five neural network models are selected to evaluate our DCT compression performance on the PASCAL VOC test dataset; the pre-trained models and testing framework are from [41], [42]. The compression strategy is based on ensuring the accuracy of the network and achieves the best compression ratio by adjusting the number of compressed layers and the Q-table values. The experimental results are shown in Table III. The combination of a sequential convolution, non-linear and pooling layer is defined as one fusion layer. This is because our accelerator can execute convolution together with non-linear and pooling layers in one data stream and only need to compress the feature maps after each fusion layer. The fully connected layer is offloaded to the CPU, so it is not within the scope of our compression method. Since the number of compressed layers of each network is different, only the compression ratios of the first ten fusion layers are given. The overall compression ratio of the whole network and the network accuracy comparison are reported in Table III. The compression ratio is defined as (20):

$$Compression\ Ratio = \frac{Data\ After\ Compression}{Origin\ Data} \quad (20)$$

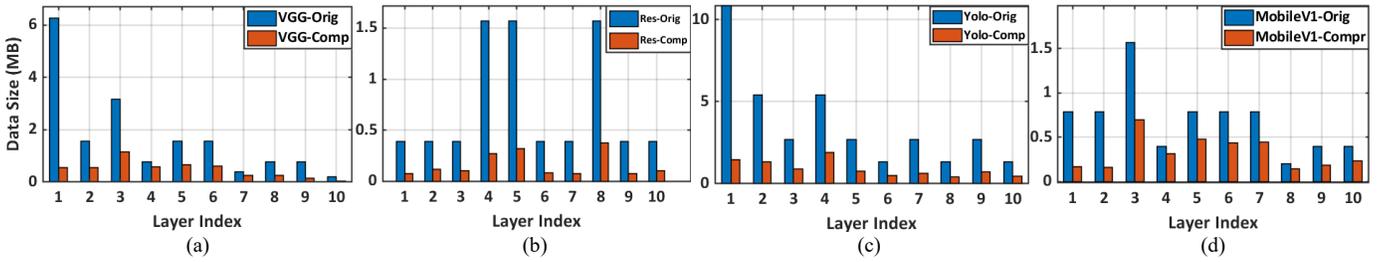

Fig.16. (a)-(d) corresponding to VGG-16-BN, ResNet-50, Yolo-v3, MobileNet-v1 original data size and compressed data size of the first 1-10 layers.

TABLE III
LAYER-BY-LAYER COMPRESSION RATIO

| Fusion[*] Layer Index | VGG-16-BN | ResNet-50 | Yolo-v3 | Mobile-Net-v1 | Mobile-Net-v2 |
|---|---|---|---|---|---|
| Fusion 1 | 8.97% | 18.99% | 13.37% | 21.05% | 27.63% |
| Fusion 2 | 34.75% | 29.36% | 24.69% | 20.68% | 31.26% |
| Fusion 3 | 37.00% | 26.47% | 32.74% | 44.38% | 88.41% |
| Fusion 4 | 72.89% | 17.39% | 35.16% | 79.85% | 48.20% |
| Fusion 5 | 42.23% | 20.09% | 28.79% | 60.28% | 77.64% |
| Fusion 6 | 38.26% | 22.02% | 36.19% | 55.67% | 56.18% |
| Fusion 7 | 67.93% | 18.63% | 23.35% | 56.76% | 66.51% |
| Fusion 8 | 31.81% | 20.93% | 31.10% | 74.82% | 68.87% |
| Fusion 9 | 18.41% | 19.66% | 27.13% | 47.26% | 57.82% |
| Fusion 10 | 27.72% | 26.14% | 34.83% | 58.30% | 61.52% |
| Overall | 30.63% | 52.51% | 65.63% | 61.02% | 71.05% |
| Origin Accuracy | 76.93% | 71.65% | 84.82% | 69.90% | 70.40% |
| Compressed Accuracy | 76.48% | 71.47% | 84.40% | 69.46% | 69.91% |

[*]The fusion layer is a combination of a convolutional layer, a batch normalization layer, an activation layer and a pooling layer. If the above operations are not included in a fusion layer, the data will be bypassed in the corresponding module.

TABLE IV
COMPARISON WITH OTHER FEATURE MAP COMPRESSION WORK

| Overall Compression Ratio | DAC'20 [16] | This Work |
|---|---|---|
| VGG-16-BN | 34.36% | 30.63% |
| ResNet-50 | 44.64% | 52.51% |
| MobileNet-v1 | N/A | 61.02% |
| MobileNet-v2 | 40.81% | 71.05% |
| On-the-fly Compression | Support | Support |
| On-chip Memory Optimization | Not Support | Support |

Without losing more than 1% accuracy, the total number of the fusion layers that can benefit from the compression ranges from 10 to 20. The first ten fusion layers have a much larger size than other layers and are often on-chip storage-limited, where off-chip memory access happens frequently. Therefore, our compression experiments mainly focus on the first ten fusion layers.

*VGG-16-BN:* By compressing the first ten fusion layers of VGG-16-BN, it achieves an overall compression ratio of 30.63%, and the overall off-chip memory access is reduced by 3.3x. The compression ratio reaches 8.97% at the first layer with the largest interlayer feature map. The original and compressed interlayer feature map data sizes are compared in Fig. 16 (a), the data size for almost all layers are less than 1 MB after compression, thus the off-chip memory access can be reduced significantly. With the compression process, the accuracy decrease is only around 0.45%.

*ResNet-50 and Yolo-v3:* ResNet-50 and Yolo-v3 are deeper with more complex network structure than VGG-16-BN. As Table III shows, the compression ratios of the first ten layers are excellent, around 20%~30% with 0.18% and 0.42% accuracy loss, respectively. As shown in Fig. 16 (b), (c), the largest interlayer feature maps in ResNet-50 are compressed to less than 0.5 MB, and most of other layers are compressed to 0.1 MB~0.2 MB, so most of the feature maps can be stored on the on-chip memory without any off-chip transmission. Although Yolo-v3 has the largest feature map data in these networks, the largest first ten fusion layers still can be compressed to between 0.5 MB and 1.5 MB.

*MobileNet-v1 and MobileNet-v2:* MobileNet-v1 and MobileNet-v2 are two lightweight networks with depth-wise convolution to achieve weight reduction. Therefore, it is difficult for further compression on these two networks. Our compression method can still achieve an overall compression ratio of 61.02% and 71.05% on these networks, reducing 1.4x-1.6x off-chip memory access with accuracy loss less than 0.5%. For the first ten layers, MobileNet-v1 and MobileNet-v2 do not perform as well as previous networks, the compression ratios are around 40%~60%. Fig. 16 (d) illustrated the interlayer data comparison of MobileNet-v1. The compression ratios of the first three layers with the largest data size are still good, and almost all the layers are compressed below 0.5 MB.

Compared to the other work [16] that implements feature map compression in Table IV, we have a better compression ratio in VGG-16-BN, but it is inferior in ResNet-50 and MobileNet-V2. Their work reported a dedicated hardware to compress data from on-chip memory to off-chip memory. Compared with them, we realized an on-the-fly compression integrated with CNN acceleration and our work compressed not only the data that needs to be transferred off-chip but also the data that is stored into the on-chip memory. Since the compression is embedded in the CNN computation flow in the accelerator, the overall on-chip memory size and off-chip access bandwidth requirement are both reduced.

Table V compares our design with other state-of-the-art works. Our accelerator can achieve 30.63%~71.05% interlayer feature map compression ratio, 403GOPS throughput, and 2.16 TOPS/W energy efficiency in TSMC 28nm technology.

TABLE V
COMPARISON WITH OTHER ACCELERATOR WORKS

| | TCASI'18 [14] | JSSC'17 [23] | JSSC'20 [28] | ISSCC'17 [24] | DATE'17 [30] | This Work |
|---|---|---|---|---|---|---|
| Technology (nm) | 65 | 65 | 65 | 28 | 28 | 28 |
| Clock Rate (MHz) | 500 | 100-250 | 200 | 200 | 700 | 700 |
| Precision (bits) | 16 | 16 | 8 | 1-16 | 16 | 16 |
| Gate Count (M) | 1.30 | 1.17 | N/A | 1.95 | 3.75 | 1.12 |
| On-chip Memory Size | 96 KB | 108 KB | 170 KB | 144 KB | 352 KB | 480 KB |
| Number of PEs | 144 | 168 | 256 | N×256 | 576 | 288 |
| Supply Voltage | 1 V | 0.82-1.17 V | 0.67-1 V | 1 V | 0.9 V | 0.72 V |
| Peak Throughput (GOPS) | 152 | 33.6-84 | 102-5638 | 102-1632 | 806 | 403 |
| Energy Efficiency (TOPS/W) | 0.434 | 0.187-0.357 | 0.411-62.1 | 0.26-10 | 1.42 | 2.16 |
| Normalized Energy Efficiency (TOPS/W)[*] | 2.34 | 1.00-1.92 | 2.21-334.66 | 0.26-10 | 1.42 | 2.16 |
| Methodology | Place & Route | Silicon | Silicon | Silicon | Place & Route | Place & Route |
| Feature Map Compression | N/A | Run Length | CSR/COO | N/A | N/A | DCT |
| Overall Compression Ratio[**] | N/A | 62.5% | 38.02% | N/A | N/A | 30.63%~71.05% |
| CNN Benchmark | VGG-16 | VGG-16 | AlexNet | VGG-16 | AlexNet | VGG-16 |
| Throughput (fps) | 4.95 | 0.7 | 2122 | 1.67 | 326.2 | 10.53 |
| Power (mW) | 350 | 236 | 20.5-248.4 | 26 | 567.5 | 186.6 |

[*] Normalized results after technology scaling [43], Normalized Energy Efficiency = Energy Efficiency $\times \kappa^2$.
[**] The Overall compression ratio of the interlayer feature map. N/A corresponding to no feature map compression implemented; 62.5% and 38.02% corresponding to the compression ratios of JSSC'17 [23] and JSSC'20 [28] on AlexNet; 30.63%~71.05% corresponding to the compression ratios of this work on VGG-16-BN, ResNet-50, MobileNet-v1, MobileNet-v2, AlexNet and Yolo-v3.

## VII. CONCLUSION

In this article, we proposed a memory-efficient CNN accelerator based on feature map compression. The proposed accelerator utilizes DCT to compress the feature maps in the frequency domain and achieves significant output feature map size reduction. An efficient hardware architecture is designed to combine compression, decompression, and CNN acceleration into one computing stream, achieving minimal processing delay for compression and high throughput. The accelerator is implemented on an FPGA and also synthesized as an ASIC IP. The results show that the accelerator can provide a higher compression ratio for state-of-the-art CNNs with negligible accuracy loss.

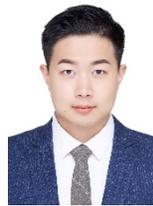

**Zhuang Shao** received the B.S. degree in telecommunications engineering from Nanjing University, Nanjing, China, in 2019, where he is currently pursuing the M.S. degree in electrical engineering. His current research interests include designs of machine-learning hardware accelerators and deep neural network compression.

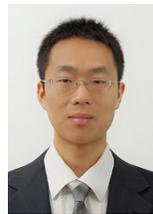

**Xiaoliang Chen** received the B.S. degree in electrical engineering from Tsinghua University, Beijing, China, in 2007, the M.S. degree in computer science from Peking University, Beijing, China, in 2010, and the Ph.D. degree in computer engineering from University of California, Irvine, USA in 2020. Since 2013, he has been with Broadcom Inc., Irvine, CA, USA, developing design flow and methodology for analog and mixed-signal circuits. His current research interests include approximate computing, machine-learning hardware accelerator and electronic design automation.

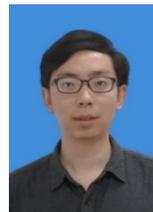

**Li Du** (M'16) received his B.S degree from Southeast University, China and his Ph.D. degree in Electrical Engineering from University of California, Los Angeles. From June 2013 to Sept 2016, he worked at Qualcomm Inc, designing mixed-signal circuits for cellular communications. From Sept. 2016 to Oct. 2018, he worked in Kneron Inc as a hardware architect research scientist, designing high-performance artificial intelligence (AI) hardware accelerator. After that, he joined XinYun Tech Inc, USA, in charge of high-speed analog circuits design for 100G/400G optical communication. Currently, he is an associate professor in the department of Electrical Science and Engineering at Nanjing University. His research includes analog sensing circuit design, in-memory computing design and high-performance AI and GPU architecture.

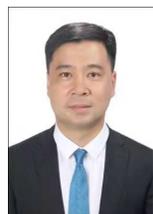

**Lei Chen** received the Ph.D. degree from the Northwestern Polytechnical University, Xi´an, China, in 2006, and completed his postdoctoral research at the China Aerospace Times Electronics Co., Ltd in 2010. Since 2007, he has been with Beijing Micro-electronics Technology Institute(BMTI), as a professor, researching high reliability integrated circuits. His current research interests are mainly about radiation hardened integrated circuits and high-performance AI.

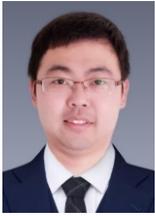

**Yuan Du** (S'14, M'19) received his B.S. degree with honor from Southeast University (SEU), Nanjing, China, in 2009, his M.S. and his Ph.D. degree both from Electrical Engineering Department, University of California, Los Angeles (UCLA), in 2012 and 2016, respectively. Since 2019, he has been with Nanjing University, Nanjing, China, as an Associate Professor. Previously, he worked for Kneron Corporation, San Diego, CA, USA ( a provider of edge AI accelerator ASICs) from 2016 to 2019, as a leading hardware architect. He has authored or coauthored more than 30 technical papers in international journals and conferences and holds over 10 US patents. His current research interests include designs of machine-learning hardware accelerators, high-speed inter-chip/intra-chip interconnects, and RFICs. He was the recipient of the Microsoft Research Asia Young Fellow (2008), Southeast University Chancellor's Award (2009), Broadcom Young Fellow (2015), and IEEE Circuits and Systems Society Darlington Best Paper Award (2021).

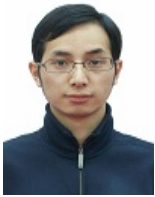

**Wei Zhuang** received his B.S and M.S degree both from Northwestern Polytechnical University (NWPU), Xi'an China in 2004 and 2007,respectively. Since Apr 2007, he has been with Beijing Microelectronics Technology Institute (BMTI), as a hardware architect research scientist, designing high-performance processor and SoC. His current research includes high-performance processor, low-power SoC and high-performance AI.

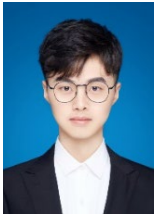

**Huadong Wei** received the B.S. degree in School of Electronic Science and Engineering from Nanjing University, Nanjing, China, in 2019, where he is currently pursuing the M.S. degree. His main research area is deep learning algorithm and hardware acceleration.

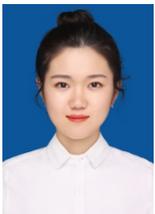

**Chenjia Xie** received the B.S. degree in electronics and information engineering from Huazhong University of science and technology, Wuhan, China, in 2020, and she is currently pursuing the M.S. degree in integrated circuit engineering. Her current research interests include designs of machine-learning hardware accelerators, compiler migration and deep neural network compression.

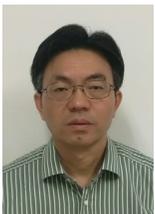

**Zhongfeng Wang** received both the B.E. and M.S. degrees in the Dept. of Automation at Tsinghua University, Beijing, China, in 1988 and 1990, respectively. He obtained the Ph.D. degree from the University of Minnesota, Minneapolis, in 2000. He has been working for Nanjing University, China, as a Distinguished Professor since 2016. Previously he worked for Broadcom Corporation, California, from 2007 to 2016 as a leading VLSI architect. Before that, he worked for Oregon State University and National Semiconductor Corporation.

Dr. Wang is a world-recognized expert on Low-Power High-Speed VLSI Design for Signal Processing Systems. He has published over 200 technical papers with multiple best paper awards received from the IEEE technical societies, among which is the VLSI Transactions Best Paper Award of 2007. He has edited one book VLSI and held more than 20 U.S. and China patents. In the current record, he has had many papers ranking among top 25 most (annually) downloaded manuscripts in IEEE Trans. on VLSI Systems. In the past, he has served as Associate Editor for IEEE Trans. on TCAS-I, T-CAS-II, and T-VLSI for many terms. He has also served as TPC member and various chairs for tens of international conferences. Moreover, he has contributed significantly to the industrial standards. So far, his technical proposals have been adopted by more than fifteen international networking standards. In 2015, he was elevated to the Fellow of IEEE for contributions to VLSI design and implementation of FEC coding. His current research interests are in the area of Optimized VLSI Design for Digital Communications and Deep Learning.